\documentclass[aps,prx,reprint,superscriptaddress,nofootinbib,longbibliography]{revtex4-2}

\usepackage[utf8]{inputenc}
\usepackage[T1]{fontenc}
\usepackage{lmodern}
\usepackage{microtype}
\usepackage{mathtools}

\usepackage{placeins}

\usepackage{amsmath}
\usepackage{amsfonts}
\usepackage{graphicx}
\usepackage{hyperref}
\hypersetup{
  colorlinks=true,
  citecolor=blue,
  linkcolor=blue,
  urlcolor=blue
}

\usepackage{mdframed}
\usepackage{comment}
\usepackage[braket,qm]{qcircuit}
\usepackage{multirow}

\newcommand{\cgate}[1]{*+<.6em>{#1} \POS ="i","i"+UR;"i"+UL **\dir{-};"i"+DL **\dir{-};"i"+DR **\dir{-};"i"+UR **\dir{-},"i" \cw}
\newcommand{\cmultigate}[2]{*+<1em,.9em>{\hphantom{#2}} \POS [0,0]="i",[0,0].[#1,0]="e",!C *{#2},"e"+UR;"e"+UL **\dir{-};"e"+DL **\dir{-};"e"+DR **\dir{-};"e"+UR **\dir{-},"i" \cw}

\begin{document}

\title{Readout Rebalancing for Near Term Quantum Computers}

\author{Rebecca Hicks}
\email{rebecca_hicks@berkeley.edu}
\affiliation{Physics Department, University of California, Berkeley, Berkeley, CA 94720, USA}

\author{Christian W. Bauer}
\email{cwbauer@lbl.gov}
\affiliation{Physics Division, Lawrence Berkeley National Laboratory, Berkeley, CA 94720, USA}

\author{Benjamin Nachman}
\email{bpnachman@lbl.gov}
\affiliation{Physics Division, Lawrence Berkeley National Laboratory, Berkeley, CA 94720, USA}

\begin{abstract}
Readout errors are a significant source of noise for near term intermediate scale quantum computers.   Mismeasuring a qubit as a $\ket{1}$ when it should be $\ket{0}$ occurs much less often than mismeasuring a qubit as a $\ket{0}$ when it should have been $\ket{1}$.  We make the simple observation that one can improve the readout fidelity of quantum computers by applying targeted $X$ gates prior to performing a measurement.  These $X$ gates are placed so that the expected number of qubits in the $\ket{1}$ state is minimized.  Classical post processing can undo the effect of the $X$ gates so that the expectation value of any observable is unchanged.  We show that the statistical uncertainty following readout error corrections is smaller when using readout rebalancing.   The statistical advantage is circuit- and computer-dependent, and is demonstrated for the $W$ state, a Grover search, and for a Gaussian state.  The benefit in statistical precision is most pronounced (and nearly a factor of two in some cases) when states with many qubits in the excited state have high probability.  
\end{abstract}

\date{\today}
\maketitle

\section{Introduction}

Quantum computers hold great promise for a variety of scientific and industrial applications.   However, existing noisy intermediate-scale quantum (NISQ) computers~\cite{Preskill2018quantumcomputingin} introduce significant errors that must be mitigated before achieving useful output.  Error mitigation on a quantum computer~\cite{0904.2557,Devitt_2013,RevModPhys.87.307,2013qec..book.....L,Nielsen:2011:QCQ:1972505} is significantly different than classical error mitigation because quantum bits (`qubits') cannot be copied~\cite{Park1970,Wootters:1982zz,DIEKS1982271}.   Full quantum error correction requires significant overhead in the additional number of qubits and gates.  This has been demonstrated for simple quantum circuits~\cite{errorcorrecting,PhysRevA.97.052313,Barends2014,Kelly2015,Linke:2017, Takita:2017, Roffe:2018, Vuillot:2018,
  Willsch:2018,Harper:2019}, but is infeasible on NISQ hardware due to limited qubit counts and circuit depths.  Instead, a variety of schemes have been proposed to mitigate -- without completely eliminating -- errors.  There are two types of errors that are targeted by these schemes: those that affect the preparation of the quantum state~\cite{Kandala:2019,Dumitrescu:2018,PhysRevX.7.021050,PhysRevLett.119.180509,PhysRevX.8.031027,PhysRevA.102.012426} and those that affect the measurement of the prepared state~\cite{bialczak_quantum_2010,neeley_generation_2010,dewes_characterization_2012,magesan_machine_2015,debnath_demonstration_2016,song_10-qubit_2017,gong_genuine_2019,wei_verifying_2020,havlicek_supervised_2019,chen_detector_2019,chen_demonstration_2019,maciejewski_mitigation_2020,urbanek_quantum_2020,nachman_unfolding_2020,hamilton_error-mitigated_2019,karalekas_quantum-classical_2020,geller_efficient_2020,geller_rigorous_2020}.  This paper focuses on the latter type -- called readout errors -- and how one can modify the quantum state prior to readout in order to reduce these errors.

Readout errors typically arise from two sources: (1) measurement times are significant in comparison to decoherence times and thus a qubit in the $\ket{1}$ state\footnote{We consider two-state systems and denote the excited state as $\ket{1}$ and the ground state as $\ket{0}$.} can decay to the $\ket{0}$ state during a measurement, and (2) probability distributions of measured physical quantities that correspond to the $\ket{0}$ and $\ket{1}$ states have overlapping support and there is a small probability of measuring the opposite value.   The first of these sources results in asymmetric errors: mismeasuring a qubit as a $\ket{1}$ when it should be $\ket{0}$ occurs much less often than mismeasuring a qubit as a $\ket{0}$ when it should have been $\ket{1}$.  We make the simple observation that one can improve the readout fidelity of quantum computers by applying targeted $X$ gates prior to performing a measurement.  These $X$ gates are placed so that the expected number of qubits in the $\ket{1}$ state is minimized.  Classical post processing can undo the effect of the $X$ gates so that the expectation value of any observable is unchanged.  This \textit{Readout Rebalancing} must be combined with additional readout corrections to unfold the migrations.   Previous studies have proposed symmetrizing the readout with targeted $X$ gates~\cite{arute2020quantum,pyquil}, but we are unaware of any proposal to introduce asymmetric rebalancing.  Symmetrizing has the advantage that it is independent of the measured state and improves the fidelity of qubits that are more often in the $\ket{1}$ state.  However, symmetrizing will necessarily reduce the fidelity for qubits that are more often in the $\ket{0}$ state (this is a state-dependent statement).

This paper is organized as follows.  Section~\ref{sec:method} introduces readout rebalancing and briefly reviews readout error mitigation.  Numerical results are presented in Sec.~\ref{sec:results} and the paper ends with conclusions and outlook in Sec.~\ref{sec:conclusions}.

\section{Method}
\label{sec:method}

To motivate the rebalancing technique, we start with an illustrative example.  Suppose that there is a two qubit system with measurement errors $\Pr(\ket{0}_i\rightarrow\ket{1}_i)=0$ and $\Pr(\ket{1}_i\rightarrow\ket{0}_i)=q_i$ for $i\in\{0,1\}$ with no nontrivial multiqubit readout errors.  When $q_i\rightarrow 0$, there are no readout errors.  For simplicity, assume that $q_i\ll 1$ so that terms of $\mathcal{O}(q_i^2)$ can be neglected.  Suppose that simple matrix inversion is used to correct readout errors and that the state is measured $N$ times.  Define $N_{\ket{ij}}$ as the number of true counts in state $\ket{ij}$ and $\hat{N}_{\ket{ij}}$ is the number of reconstructed counts in state $\ket{ij}$ following readout error corrections.  The readout corrections are estimated by inverting the $4\times 4$ matrix encoding the transition probabilities between any possible true state and any possible observed state.  By construction, $\mathbb{E}[\hat{N}_{\ket{ij}}]=N_{\ket{ij}}$. One can show that the variance of the counts after readout correction are given by (for details, see Appendix~\ref{AppendixA})
\begin{align}
\text{Var}[\hat{N}_{\ket{ij}}] = N_{\ket{ij}} \left( 1 - \frac{N_{\ket{ij}}}{N} \right) + \Delta \text{Var}[\hat{N}_{\ket{ij}}]
\end{align}
with
\begin{align}
\label{eq:simple1}
\Delta\text{Var}[\hat{N}_{\ket{00}}]&= q_0 N_{\ket{10}} + q_1 N_{\ket{01}}+\mathcal{O}(q^2) 
\nonumber\\
\Delta\text{Var}[\hat{N}_{\ket{11}}]&=\left(q_0+q_1\right)N_{\ket{11}}+\mathcal{O}(q^2)\\\nonumber
\end{align}
In particular, if $N=N_{\ket{11}}$ (i.e. the other states have zero true counts), one finds $\text{Var}[\hat{N}_{\ket{00}}] = 0$, while $\text{Var}[\hat{N}_{\ket{11}}] \neq 0$, for non-vanishing $q_i$. On the other hand, if $N=N_{\ket{00}}$, both $\text{Var}[\hat{N}_{\ket{00}}] = 0$ and $\text{Var}[\hat{N}_{\ket{11}}] = 0$ vanish.   This suggests that if one is trying to measure a state dominated by $\ket{11}$, it would be more effective to first invert $0\leftrightarrow 1$, perform the measurement, and then swap back the classical bits afterward.

\begin{figure}

\[
\Qcircuit @C=1em @R=0.8em {
    \lstick{\ket{0}}  & \multigate{5}{U_\text{circuit}} & \qw &\qw& \meter & \cw \cw[1] &\cmultigate{5}{R^{-1}} &\cw &\cw &\cw\\
    \lstick{\ket{0}}  & \ghost{U_\text{circuit}} & \qw &\qw&\meter & \cw \cw[1] & \pureghost{R^{-1}} & \cw  & \cw&\cw \\
    \lstick{\ket{0}}  & \ghost{U_\text{circuit}} & \qw &\qw&\meter& \cw \cw[1] & \pureghost{R^{-1}} & \cw  & \cw&\cw \\
    \lstick{\ket{0}}  & \ghost{U_\text{circuit}} & \qw &\qw& \meter& \cw \cw[1] & \pureghost{R^{-1}} & \cw  & \cw&\cw \\
    \lstick{\ket{0}}  & \ghost{U_\text{circuit}} & \qw &\qw&\meter& \cw \cw[1] & \pureghost{R^{-1}} & \cw  &\cw&\cw \\
    \lstick{\ket{0}}  & \ghost{U_\text{circuit}} & \qw &\qw&\meter&\cw \cw[1] & \pureghost{R^{-1}} & \cw  & \cw&\cw\\
    &&&&&\\
    &&&&\big\downarrow&&\\
    &&&&&\\
    \lstick{\ket{0}}  & \multigate{5}{U_\text{circuit}} & \qw & \gate{X}& \meter & \cw \cw[1] &\cmultigate{5}{R^{-1}} &\cw &\cgate{X} &\cw\\
    \lstick{\ket{0}}  & \ghost{U_\text{circuit}} & \qw &\qw&\meter & \cw \cw[1] & \pureghost{R^{-1}} & \cw  & \cw&\cw \\
    \lstick{\ket{0}}  & \ghost{U_\text{circuit}} & \qw &\qw&\meter& \cw \cw[1] & \pureghost{R^{-1}} & \cw  & \cw&\cw \\
    \lstick{\ket{0}}  & \ghost{U_\text{circuit}} & \qw &\qw& \meter& \cw \cw[1] & \pureghost{R^{-1}} & \cw  & \cw&\cw \\
    \lstick{\ket{0}}  & \ghost{U_\text{circuit}} & \qw &\gate{X}&\meter& \cw \cw[1] & \pureghost{R^{-1}} & \cw  &\cgate{X}&\cw \\
    \lstick{\ket{0}}  & \ghost{U_\text{circuit}} & \qw &\qw&\meter&\cw \cw[1] & \pureghost{R^{-1}} & \cw  & \cw&\cw
     }
\]

\caption{An illustration of the Readout Rebalancing protocol.  Here $U_\text{circuit}$ represents the state preparation that must happen for each measurement, and the readout error mitigation represented by an inverted response matrix $R^{-1}$ (in practice, a more sophisticated readout error mitigation scheme may be used) is performed on an ensemble of measured states. From the measured values of the qubits of the first circuit which uses a small fraction of the total number of runs one then determines which qubits have $\langle q_i \rangle > 0.5$ and should therefore be flipped (the first and fifth in our example). One then runs the remaining large fraction of the runs on the modified second circuit.}
\label{fig:schematic}
\end{figure}
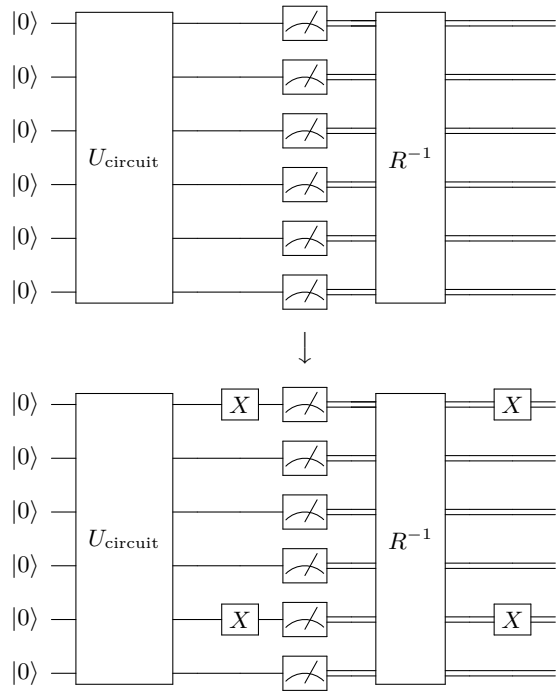

The readout rebalancing protocol is illustrated in Fig.~\ref{fig:schematic}.  First, the probability mass function over states $p(x)$ is estimated using a small fraction of the total number of intended measurements.  Then, a rule is used to determine which qubits should be flipped prior to being measured, with the goal of switching those qubits that are predominantly in the $\ket{1}$ state.  There are multiple possible rules, and in this paper we use a simple and effective approach. In this approach one first computes $\langle q_i\rangle$, the average value for each qubit $i$.  If this value is greater than 0.5, the qubit $i$ is set to be flipped and otherwise it is untouched.  A modified circuit is then prepared where single qubit $X$ gates are applied at the end of the circuit to the qubits set to be flipped.  Using this modified circuit one then performs the intended measurements.  These data are corrected for readout errors (more on this below) and then post-processed with classical $X$ gates to undo the quantum $X$ gates.

As stated earlier, the reason that the readout rebalancing protocol is expected to be effective is that errors are asymmetric.  While we are not aware of other proposals for biased (circuit-specific) rebalancing, Google AI Quantum~\cite{arute2020quantum} (through its software \textsc{Cirq}~\cite{cirq}) and Rigetti (through its software \texttt{pyQuil}~\cite{pyquil}) have proposed symmetric rebalancing whereby the result from a circuit is averaged with a version of its complement that uses some pre-determined (non-state-specific) set of $X$ gates.  This approach improves the fidelity of qubits that are mostly in the one state, but it degrades the performance of measuring states where a qubit is mostly in the zero state.  We will compare to a version of symmeterized readout where measurements are averaged using a nominal circuit and a circuit with all $X$ gates.

There are many options for readout error corrections, indicated by $R^{-1}$ in Fig.~\ref{fig:schematic}.  Let $m$ and $t$ represent the raw and true probability mass functions of the state, respectively.  Furthermore, $R_{ij}=\Pr(m=i|t=j)$ is an estimate of the response matrix.  The simplest readout error correction approach is simply $\hat{t}_\text{matrix}=R^{-1}t$.  For a variety of reasons, this may be suboptimal and so multiple alternative methods have been proposed~\cite{1904.11935,1907.08518,nachman_unfolding_2020}.  In this paper, we will use Iterative Bayesian Unfolding (IBU)~\cite{DAgostini:1994fjx,1974AJ.....79..745L,Richardson:72} as described in the context of quantum computing in Ref.~\cite{nachman_unfolding_2020}.  This iterative procedure starts with $\hat{t}^0_i=\Pr(m=i)=1/2^{n_\text{qubits}}$ and then 
\begin{align}
    \hat{t}_{i}^{n+1}&=\sum_j\Pr(t=i|m=j)\times m_j\\
    &=\sum_h\frac{R_{ji}\hat{t}_i^n}{\sum_k R_{jk}\hat{t}_k^n}\times m_j.
\end{align}
We will use $100$ iterations and label $\hat{t}\equiv\hat{t}^{100}$, but the results do not depend strongly on this number.

The next session will illustrate readout rebalancing for several example states by combining it with IBU.

\section{Results}
\label{sec:results}

\begin{figure}[h!]
\centering
\includegraphics[width=0.4\textwidth]{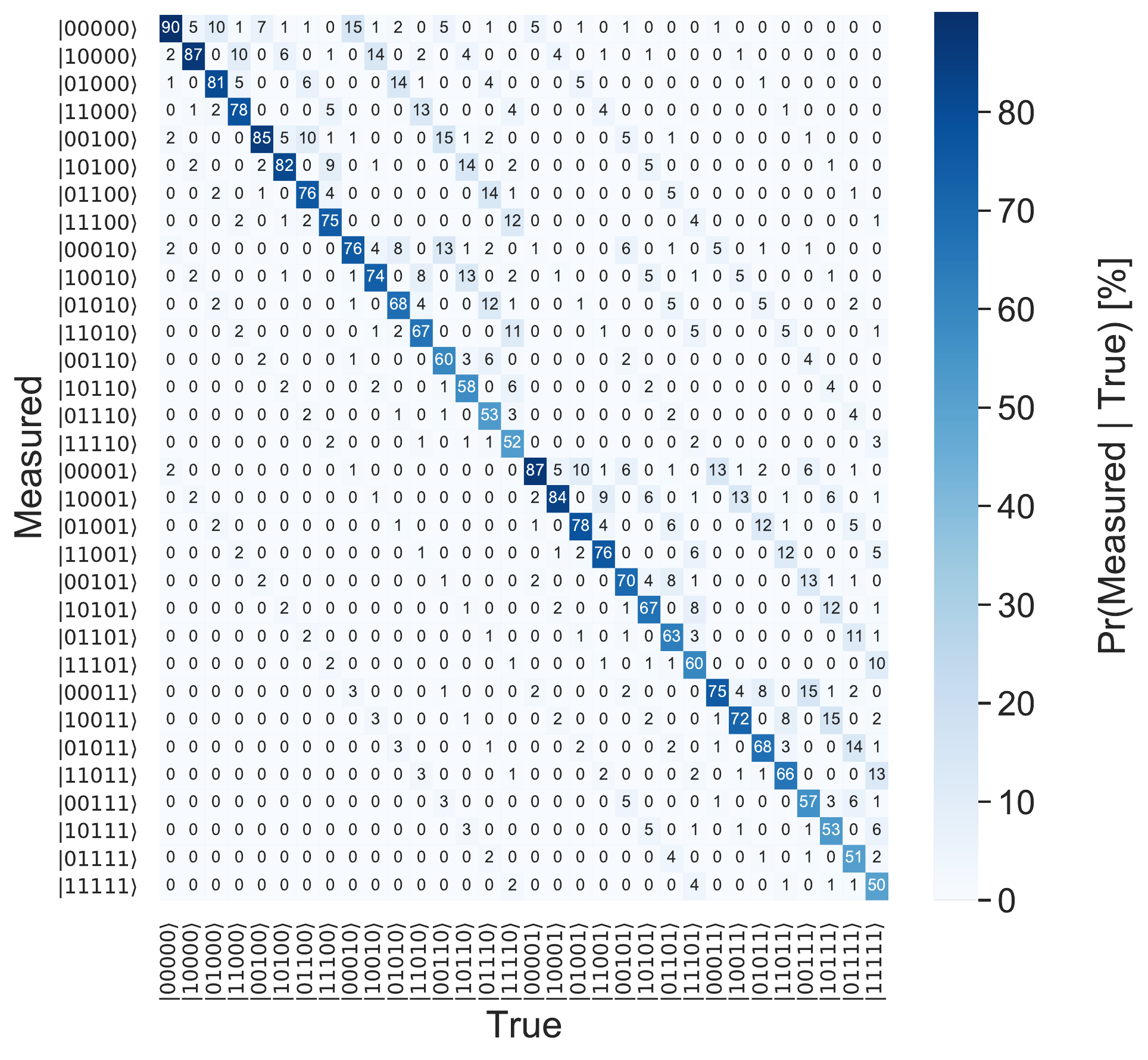}\\
\includegraphics[width=0.4\textwidth]{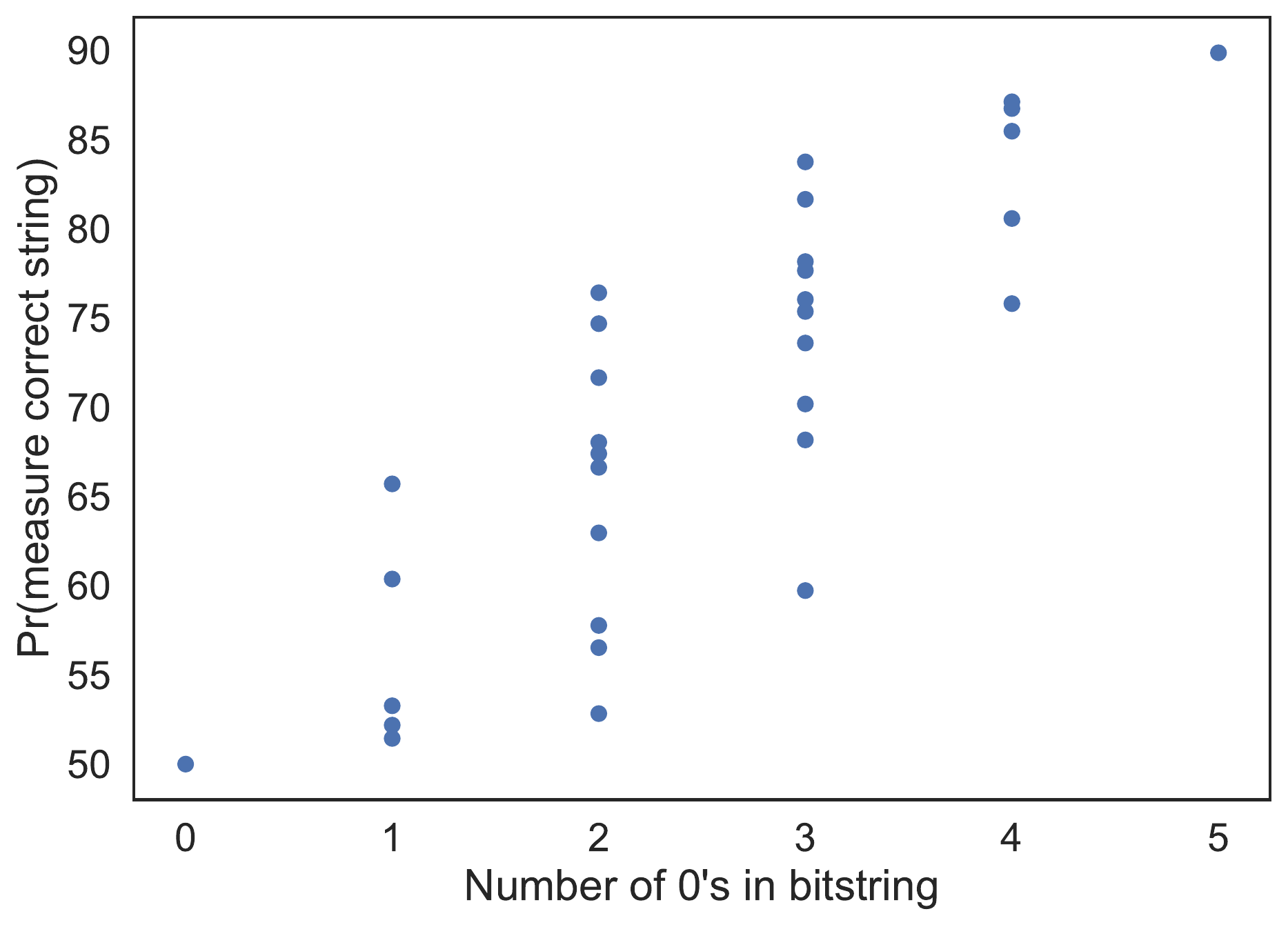}
\caption{Top: An example response matrix for the IBM Q Tokyo machine with five qubits. Bottom: the probability to measure the correct string (diagonal elements of the top plot) as a function of the number of 0's in the bitstring.}
\label{fig:response}
\end{figure}

To demonstrate readout error corrections, we utilize the software package \texttt{Qiskit} by IBM~\cite{qiskit}.  Readout errors from the IBM Q Tokyo machine (see Ref.~\cite{unfolding}) are imported for illustrating the impact of readout rebalancing.  The corresponding response matrix is presented in Fig.~\ref{fig:response}.  Most of the probability mass in the response matrix is along the diagonal, which represents the probability for a particular state to be correctly measured.  However, there are significant off-diagonal terms, which are more pronounced in the lower right part of the matrix.  To illustrate this feature, the bottom plot in Fig.~\ref{fig:response} shows the probability for a state to be correctly measured organized by the number of $0$'s in the state bitstring.  As advertised earlier, more 1's in the bitstring correlates with a lower probability of being measured correctly.

\begin{figure}[h!]
\centering
\includegraphics[width=0.4\textwidth]{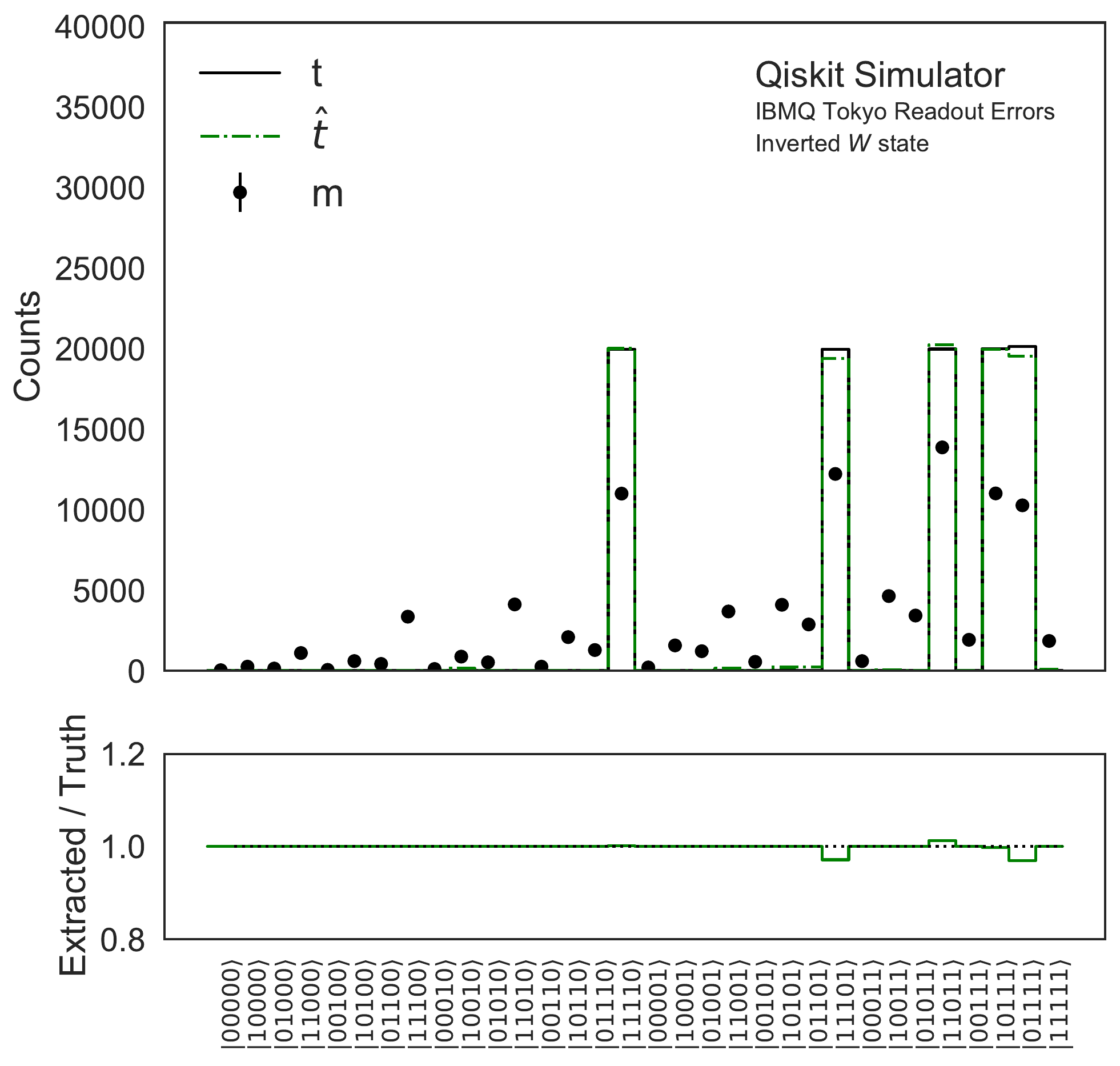}\\
\includegraphics[width=0.4\textwidth]{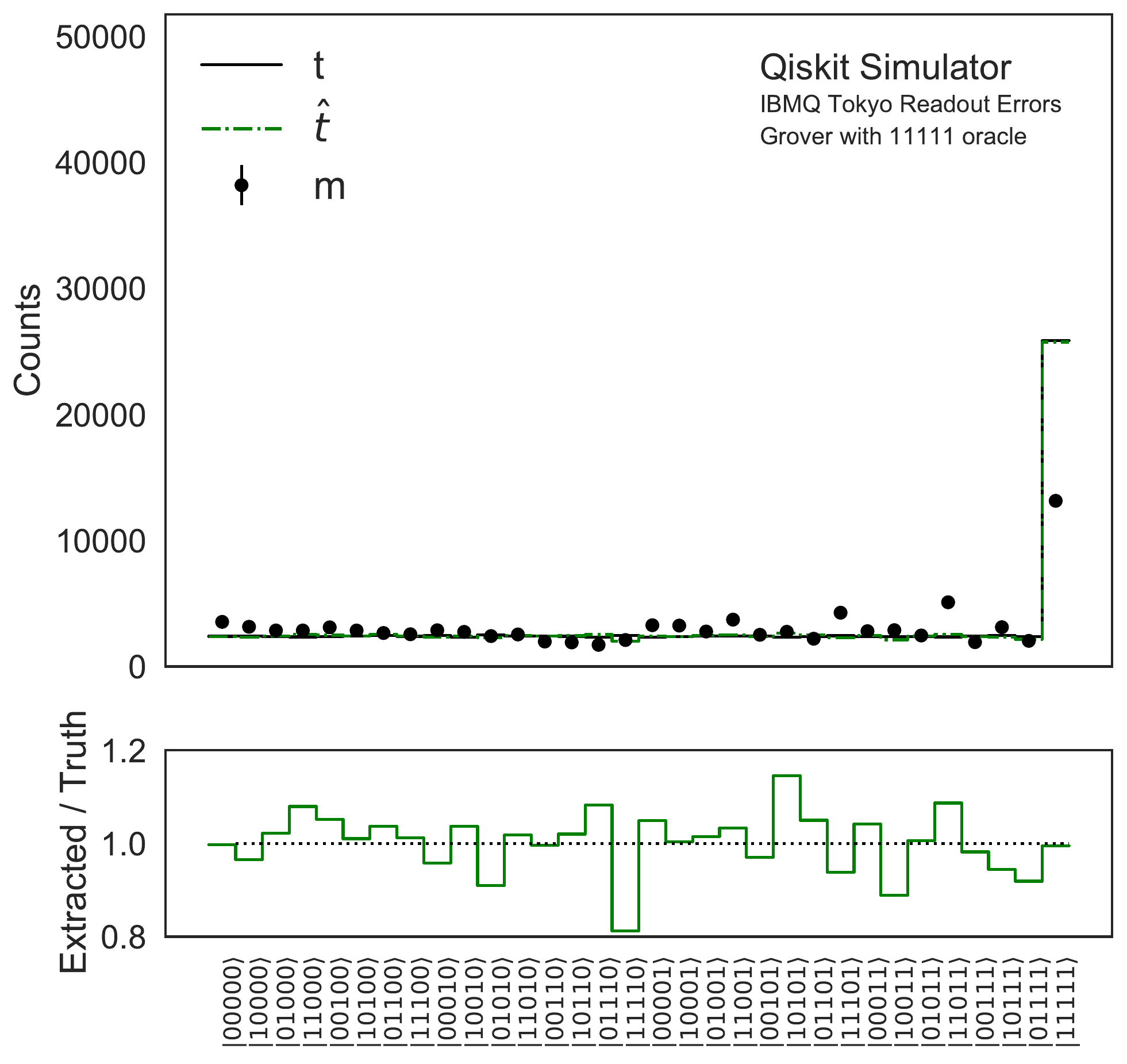}\\
\caption{The inverted $W$ state (top) and Grover state with $11111$ oracle (bottom) for the truth counts ($t$), the raw counts ($m$) and the readout corrected counts ($\hat{t}$).}
\label{fig:invertedWstate}
\end{figure}
The first example to demonstrate readout rebalancing uses a circuit  $U_{\rm circuit}$ producing the inverted $W$ state:
\begin{align}
\label{eq:invertedW}
\psi_{W,I}=\frac{1}{\sqrt{5}}\left(\ket{01111}+\ket{10111}+\cdots+\ket{11110}\right)\,.
\end{align}
Measured counts from sampling $\psi_{W,I}$ are shown in the top plot of Fig.~\ref{fig:invertedWstate}.  The true distribution has five spikes corresponding to the five stats with non-zero amplitude in Eq.~\ref{eq:invertedW}. The raw data have non-trivial probability mass for other states due to readout errors (the off-diagonal elements in Fig.~\ref{fig:response}).  Readout error corrections successfully morph the raw data $m$ into $\hat{t}$ which closely resembles the true distribution $t$.

The impact of of our method can be seen by comparing the statistical uncertainty of a particular expectation value obtained with and without readout rebalancing.  Statistical uncertainties are determined by repeating the above procedure many (1000) times and then computing the mean and standard deviation.  To be concrete, we choose the average value of the integer obtained from the bitstring
\begin{align}
\label{eq:o}
    \langle\mathcal{O}\rangle=\frac{1}{N}\sum_s n_s (s_0+2s_1+4s_2+8s_2+16 s_3+32 s_4)\,,
\end{align}
 as a representative observable. Here
$s_i\in\{0,1\}$ is the value of the $i^\text{th}$ qubit in state $s$, $n_s$ is the number of times the state is measured to be in state $s$, and $N=\sum_s n_s$ is the total number of measurements in one iteration of the procedure.  For the inverted $W$ state, the exact value of this observable is given by $\langle\mathcal{O}\rangle= 124 / 4 = 24.8$. Computing the average value of $\langle\mathcal{O}\rangle$ from repeating the procedure 1000 times reproduces this result with and without readout symmetrization and readout rebalancing.  This is expected, since the post-processing step should not affect the central value.  In contrast, the standard deviation across the 1000 runs of the procedure depends on the level of fidelity improvement one applies. Without any fidelity improvement, the standard deviation is given by $0.0232\pm 0.0005$, with symmetrized readout it is $0.0214\pm0.0005$, while for readout rebalancing one finds a standard deviation of $0.0189\pm0.004$.  This means that one can use fewer measurements with readout rebalancing to achieve the same statistical precision as the nominal approach or for the same number of measurements, the statistical uncertainty is smaller with readout rebalancing included.  In particular, reducing the statistical uncertainty by 20\% is equivalent to having $50\%$ more measurements (since $\sqrt{1.5}\sim 1.2$).

As a second example  we use Grover's algorithm~\cite{10.1145/237814.237866}, shown in the bottom plot of Fig.~\ref{fig:invertedWstate}. In general, given an oracle, Grover's algorithm is able to find the inputs that produce a particular output value. In our case, the oracle is the Boolean function $(x_0\vee x_1\vee x_2 \vee x_3\vee x_4)$, where the input to $x_i$ would be the $i^\text{th}$ bit of our bitstring. For this function to equal 1 our desired input would be the state $\ket{11111}$, so instead of Eq.~\eqref{eq:o}, we use the counts in the $\ket{11111}$ state as our observable for $10^5$ measurements.  As with the previous example, all three approaches achieve the consistent central values ($2.57\times 10^4$).  However, the standard deviations are $210\pm 4$ (default), $185\pm4$ (symmetrized readout), and $160\pm4$ (readout rebalancing).  This 30\% improvement in precision is equivalent to a 70\% larger number of measurements (since $\sqrt{1.7}\sim 1.3$).

\begin{figure}[h!]
\centering
\includegraphics[width=0.4\textwidth]{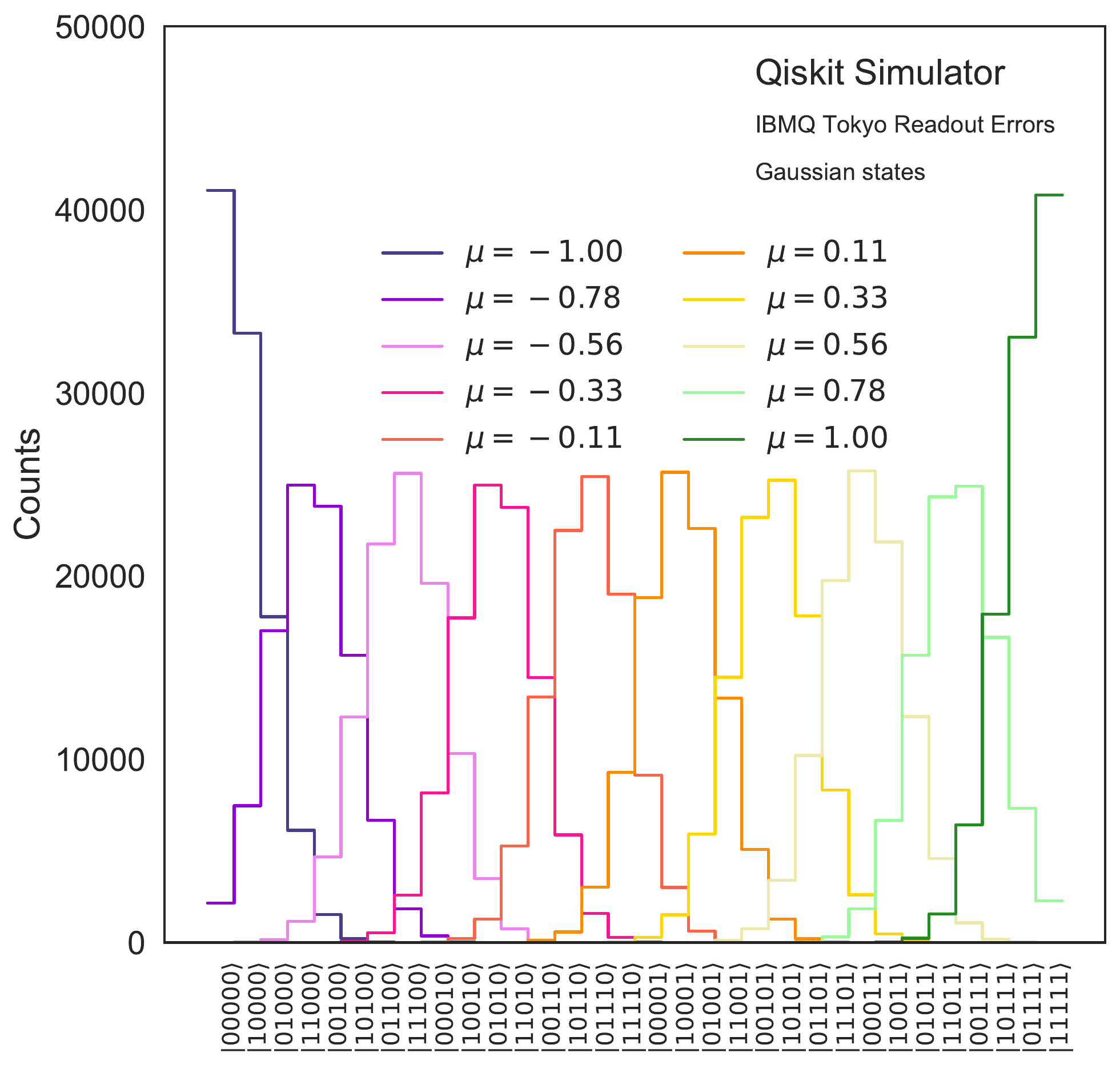}
\includegraphics[width=0.4\textwidth]{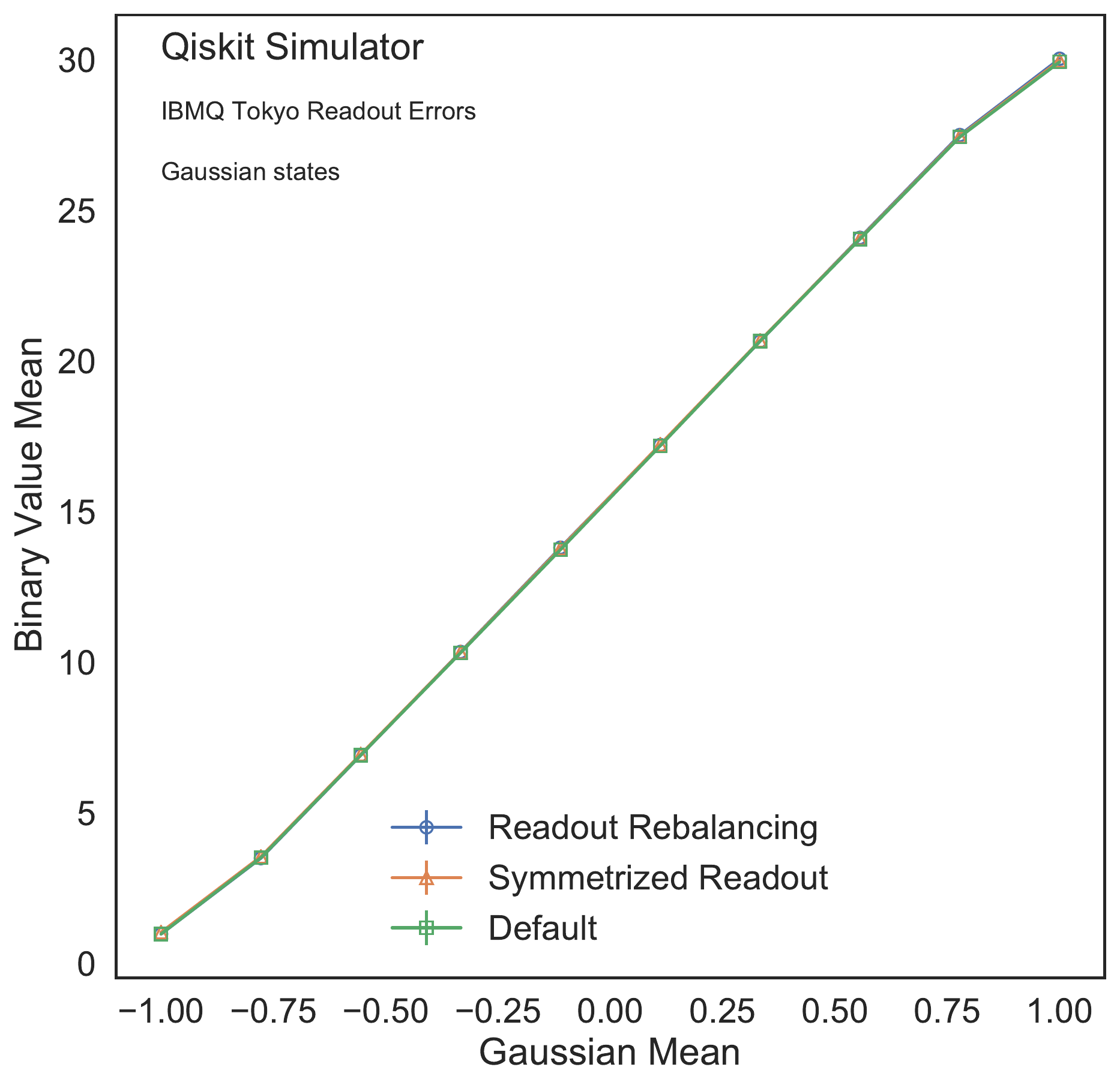}
\includegraphics[width=0.4\textwidth]{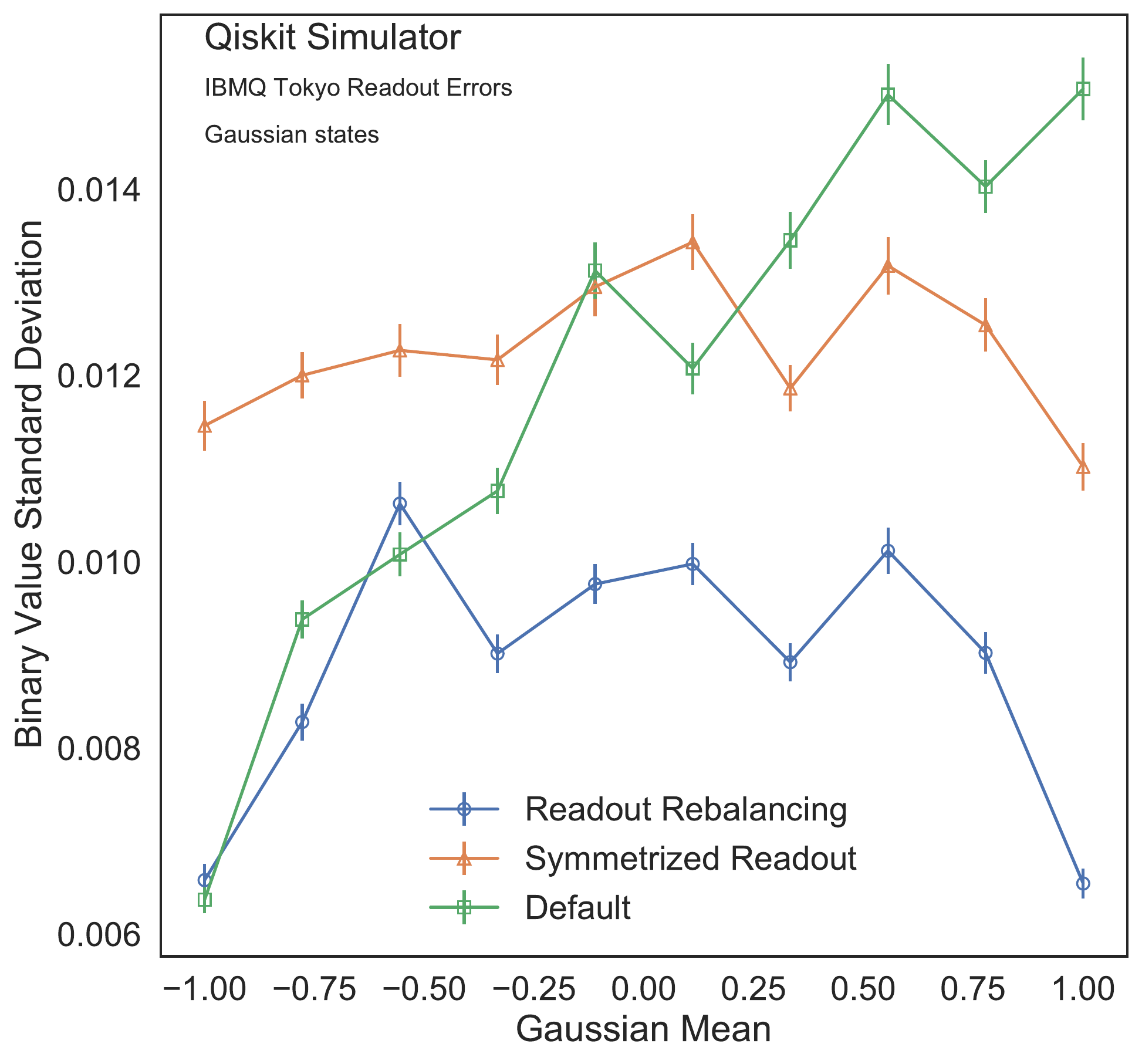}
\caption{Top: the true counts for the truncated Gaussian states with fixed variance and shifted means.  Middle (right): the average value (standard deviation) of the state when converting the bitstring to base 10 as a function of the Gaussian mean.  Error bars represent statistical uncertainties.}
\label{fig:Gaussian}
\end{figure}
A third example is a one-dimensional Gaussian state, which arises in the context of a 0 + 1 dimensional non-interacting scalar quantum field theory (as the ground state of the Harmonic Oscillater)~\cite{Jordan:2017lea,Jordan:2011ci,Jordan:2011ne,Jordan:2014tma,10.5555/3179430.3179434,PhysRevLett.121.110504,Macridin:2018oli,Klco:2018zqz}. In order to illustrate the impact of readout rebalancing, a Gaussian random variable with mean $\mu$ and standard deviation 0.1 is digitized with 5 bits where $\ket{00000}\mapsto -1$ and $\ket{11111}\mapsto 1$, where $\mu$ ranges from $-1$ to $1$.  Probability mass functions of these Gaussian states are presented in the top plot of Fig.~\ref{fig:Gaussian}.

\begin{table}[h!]
\label{tab}
\begin{tabular}{l|c|c}
& Symmetrized & {\bf Rebalanced}\\\hline
Inverted W & $(85 \pm 6)$\% & {\bf $\mathbf{(66 \pm 5)}$\%} \\
Grover & $(78 \pm 5)$\% & {\bf $\mathbf{(58 \pm 5)}$\%} \\
Gaussian $(\mu = -0.11)$ & $(98 \pm 6)$\% & {\bf $\mathbf{(56 \pm 5)}$\%} \\
Gaussian $(\mu = 0.78)$ & $(80 \pm 6)$\% & {\bf $\mathbf{(41 \pm 4)}$\%}
\end{tabular}
\caption{Table summarizing the three examples given in the text. We show the fraction of events needed to obin the same statistical power as for readout correction without any readout fidelity improvements.}
\end{table}
The mean and standard deviation of $\langle\mathcal{O}\rangle$ are shown in the middle and bottom plots of Fig.~\ref{fig:Gaussian}, respectively.  As expected, $\langle\mathcal{O}\rangle$ increases monotonically with $\mu$ and is the same with and without readout rebalancing.  However, due to the larger number of 1's on the right side of the digitized domain of the Gaussian, readout rebalancing results in a smaller statistical uncertainty than the nominal approach.  Readout rebalancing results in a relatively constant statistical precision as a function of $\mu$, with a slight increase in the middle of the domain due to a balanced number of 0's and 1's.  With readout rebalancing, the right side of the domain is equivalent to the left side.  For $\mu$ close to one, the improvement in the statistical uncertainty is nearly a factor of two.

We summarize the results of the three examples in Table~\ref{tab}. In this table we present the results by showing the fraction of events that are required to achieve the same statistical power compared with the case where no rebalancing is performed before readout error correction. One can see that the Rebalanced approach propsed in this work outperforms the Symmetrized result, and that by performing readout rebalancing one can save about a factor of 2 in the number of measurements required.

\section{Conclusions}
\label{sec:conclusions}

We have introduced a modification to readout error mitigation that can be combined with other approaches to readout error corrections.  While the benefit of this readout rebalancing scheme are circuit-dependent and may be modest, there is minimal computational cost and when states with many $1$'s are frequent, the gain in statistical precision can be significant.

\section*{Code and Data}

The code for this paper can be found at \url{https://github.com/LBNL-HEP-QIS/ReadoutRebalancing}.  Quantum computer data are available upon request.

\begin{acknowledgments}

We would like to thank Miro Urbanek and Bert de Jong for useful discussions and feedback on the manuscript.  This work is supported by the U.S. Department of Energy, Office of Science under contract DE-AC02-05CH11231. In particular, support comes from Quantum Information Science Enabled Discovery (QuantISED) for High Energy Physics (KA2401032) and the Office of Advanced Scientific Computing Research (ASCR) through the Accelerated Research for Quantum Computing Program.   This research used resources of the Oak Ridge Leadership Computing Facility, which is a DOE Office of Science User Facility supported under Contract DE-AC05-00OR22725.

\end{acknowledgments}

\appendix

\section{Detailed Derivation of Analytic Example}
\label{AppendixA}

Given a state containing 2 qubits there are 4 possible states $\ket{00}$, $\ket{01}$, $\ket{10}$, and $\ket{11}$. Prior to readout error corrections (PRC), the measured count for the state $\ket{ij}$ is given by $\hat{N}_{\ket{ij}}^\text{PRC}$. The expectation values $\mathbb{E}[\hat{N}_{\ket{ij}}^\text{PRC}] = \hat{N}_{\ket{ij}}^\text{PRC} / N$ of those counts are related to the expectation values of the true counts via
\begin{align}
\label{eq:prc1}
    \mathbb{E}[\hat{N}_{\ket{00}}^\text{PRC}]&=\mathbb{E}[N_{\ket{00}}]+q_0\mathbb{E}[N_{\ket{10}}]+q_1\mathbb{E}[N_{\ket{01}}]\\\label{eq:prc2}
    \mathbb{E}[\hat{N}_{\ket{01}}^\text{PRC}]&=(1-q_1)\mathbb{E}[N_{\ket{01}}]+q_0\mathbb{E}[N_{\ket{11}}]\\\label{eq:prc3}
    \mathbb{E}[\hat{N}_{\ket{10}}^\text{PRC}]&=(1-q_0)\mathbb{E}[N_{\ket{10}}]+q_1\mathbb{E}[N_{\ket{11}}]\\\label{eq:prc4}
    \mathbb{E}[\hat{N}_{\ket{11}}^\text{PRC}]&=(1-q_0-q_1)\mathbb{E}[N_{\ket{11}}]+\mathcal{O}(q^2)\,.
\end{align}

One can therefore obtain the expectation values of the true counts by computing the reconstructed counts from the measured counts as
\begin{align}
    \hat{N}_{\ket{00}}&=\hat{N}_{\ket{00}}^\text{PRC}-q_1\hat{N}_{\ket{01}}^\text{PRC}-q_0\hat{N}_{\ket{10}}^\text{PRC}+\mathcal{O}(q^2)\\
    \hat{N}_{\ket{01}}&=\hat{N}_{\ket{01}}^\text{PRC}(1+q_1) - q_0 \hat{N}_{\ket{11}}^\text{PRC}+\mathcal{O}(q^2)\\
    \hat{N}_{\ket{10}}&=\hat{N}_{\ket{10}}^\text{PRC}(1+q_0) - q_1 \hat{N}_{\ket{11}}^\text{PRC}+\mathcal{O}(q^2)\\
    \hat{N}_{\ket{11}}&=(1+q_0+q_1)\hat{N}_{\ket{11}}^\text{PRC}+\mathcal{O}(q^2) \,.   
\end{align}
and requiring $\mathbb{E}[\hat{N}_{\ket{ij}}]=\mathbb{E}[N_{\ket{ij}}]$.

The variance of $\hat N_{\ket{ij}}$ determines how many overall runs are required to obtain $\mathbb{E}[{N}_{\ket{ij}}]$ with a given accuracy. One way of viewing $\hat{N}_{\ket{ij}}^\text{PRC}$ is as a multinomial random variable with total number $N$ and probabilities given by Eq.~\eqref{eq:prc1}-\eqref{eq:prc4} via $p_{ij}=\mathbb{E}[\hat{N}_{\ket{ij}}^\text{PRC}]/N$.  Therefore,
\begin{align}
    \text{Var}[\hat{N}_{\ket{ij}}^\text{PRC}]&=p_{ij}(1-p_{ij})N\\
    \text{Cov}[\hat{N}_{\ket{ij}}^\text{PRC},\hat{N}_{\ket{i'j'}}^\text{PRC}]&=-p_{ij}p_{i'j'}N
\end{align}

Using the fact that
\begin{align}\nonumber
    \text{Var}[aX+bY]&=a^2\text{Var}(X)+b^2\text{Var}(Y)\\
    &\qquad\quad+2ab\,\text{Cov}(X,Y),
\end{align}
one can derive
\begin{align}
\text{Var}[\hat{N}_{\ket{ij}}] = N_{\ket{ij}} \left( 1 - \frac{N_{\ket{ij}}}{N} \right) + \Delta \text{Var}[\hat{N}_{\ket{ij}}]
\end{align}
with
\begin{align}
\label{A10}
\Delta\text{Var}[\hat{N}_{\ket{00}}]&= q_0 N_{\ket{10}} + q_1 N_{\ket{01}} \\
\Delta\text{Var}[\hat{N}_{\ket{01}}]&= q_0 N_{\ket{11}} + q_1 N_{\ket{01}}\\
\Delta\text{Var}[\hat{N}_{\ket{10}}]&= q_1 N_{\ket{11}} + q_1 N_{\ket{10}}\\
\Delta\text{Var}[\hat{N}_{\ket{11}}]&=\left(q_0+q_1\right) N_{\ket{11}}
\label{A13}
\,,\end{align}
where Eqs.~\eqref{A10} - \eqref{A13} have all been expanded to linear order in the $q_i$, and therefore have corrections of order $q_i^2$.

\bibliography{myrefs}

\end{document}